\documentclass[11pt]{article}
\usepackage{blois2002,epsfig}
\def\Journal#1#2#3#4{{#1} {\bf #2}, #3 (#4)}

\def\NCA{{\em Nuovo Cimento}}

\def\PLB{{\em Phys. Lett.}  B}
\def\PRL{{\em Phys. Rev. Lett.}}
\def\PRD{{\em Phys. Rev.} D}
\def\ZPC{{\em Z. Phys.} C}
\def\PTP{{\em Prog. Theor. Phys. }}

\def\babar{\mbox{\sl B\hspace{-0.1cm} {\small\sl A}\hspace{-0.1cm} 
\sl B\hspace{-0.1cm} {\small\sl A\hspace{-0.05cm}R}}}
\def\bm{{\it B}}

\def\be{\begin{equation}}
\def\ee{\end{equation}}
\def\bea{\begin{eqnarray}}
\def\eea{\end{eqnarray}}

\begin{document}
\vspace*{4cm}
\title{B Physics at Hadron Colliders}

\author{ D. Bortoletto }

\address{Department of Physics, Purdue University, 525 Northwestern Avenue,\\
West Lafayette, Indiana 47907, U.S.A.}

\maketitle\abstracts{
Hadron colliders have played, and will continue to play, a crucial role in the
understanding of CP violation.  Their impact is expected to be especially important 
in the study of CP asymmetries in rare decays and the $B_s$ meson. The 
experimental challenge and reach of current and future experiments is discussed.
}

\section{Introduction}

Almost forty years have past since the first observation of CP violation in the kaon 
system. Recently, CP violation has been clearly established in the B system 
with the measurement of sin 2$\beta$ at the $e^+e^-$ \bm-factories. The average value of 
$\sin 2\beta=0.734 \pm 0.054 $ measured
by the Belle and BaBar experiments~\cite{sin2betaexp} is in excellent agreement 
with the expectation of the standard model fits\cite{sin2betafit} which are based 
on the Cabibbo, Kobayashi and Maskawa (CKM) ansatz~\cite{ckm}. In fact, already 
in 1998 several groups predicted a value of $\sin 2\beta=0.72 \pm 0.02$ which was 
remarkably close to the current world average\cite{oldfit}. 

Despite the great success of the CKM description of 
CP violation it is clear that we have not yet answered the 
fundamental questions about the origin of the hierarchical pattern of the CKM 
matrix elements that we are now confirming experimentally. A major goal of
present and future experiments is 
to fully tests the CKM ansatz. Measurements at hadron colliders have the potential 
to facilitate this task by providing a unique laboratory to 
study CP violation in the $B_d$ and $B_s$ system, 
mixing of the $B_s$ meson, and rare $B$ decays.  
Many studies on the impact of hadronic machines in $B$ physics
have recently taken place
and were a source for this talk.\cite{{LHC},{TEV}}

\subsection{CP Violation in the Standard Model and the CKM Matrix}

The CKM matrix, $V_{CKM}$, is a unitary matrix \cite{ckm} that transforms the mass 
eigenstates to the weak eigenstates. The matrix can be expressed in terms of four 
independent phases which are not predicted in the standard model but have to be determined 
experimentally. In the Wolfenstein parameterization \cite{wolf}
the matrix can be written in terms of $\lambda$, $A$, $\rho$ and $\eta$ as:  
\begin{eqnarray*}
\lefteqn
{V_{CKM}=\left(\begin{array}{ccc}
V_{ud} & V_{us} & V_{ub} \\ 
V_{cd} & V_{cs} & V_{cb} \\ 
V_{td} & V_{ts} & V_{tb} \\ 
\end{array}\right) 
 \approx  }  \\
& & \left(\begin{array}{ccc}
1-\frac{\lambda^2}{2} & \lambda & A\lambda^3(\rho-i\eta (1- \frac{\lambda^2}{2}) ) \\ 
-\lambda & 1-\frac{\lambda^2}{2}-i \eta A^2 \lambda^4 & 
 A\lambda^2(1+i\eta\lambda^2) \\ 
 A\lambda^3(1-\rho-i\eta ) & -A\lambda^2  &1 \\ 
\end{array}\right)
\end{eqnarray*}
The expression is accurate to order $\lambda^3$ in the real part and 
$\lambda^5$ in the imaginary part. 
The sine of the Cabibbo angle, $\lambda$, is measured in semileptonic
kaon decays, $\lambda=|V_{us}|=0.2196\pm 0.0026$, and plays the role
of an expansion parameter. $A$ can be determined in
$b\rightarrow c$ decays since
$A \approx V_{cb}/\lambda^2$ and $V_{cb}=(41.2 \pm 2.0) \times 10^{-3}$. 
Only $\lambda$ and $A$ are measured
precisely. The measurements of $\sin 2\beta$, $\epsilon$ in Kaon decays and 
$B$ mixing constraint $\eta$ and $\rho$ 
with some accuracy and the PDG 2002 fit\cite {PDG}
yields $\bar{\rho}$=$\rho(1-\frac{\lambda^2}{2})$= $0.22 \pm 0.10$ and $\bar{\eta}$=
$\eta(1-\frac{\lambda^2}{2})$=$0.35 \pm 0.05$. The parameter $\eta$ represents the CP-violating 
phase and must be 
different from zero to accommodate CP violation in the standard model.

\begin{figure}
%\rule{5cm}{0.2mm}\hfill\rule{5cm}{0.2mm}
%\vskip 2.5cm
%\rule{5cm}{0.2mm}\hfill\rule{5cm}{0.2mm}
\centerline{\psfig{figure=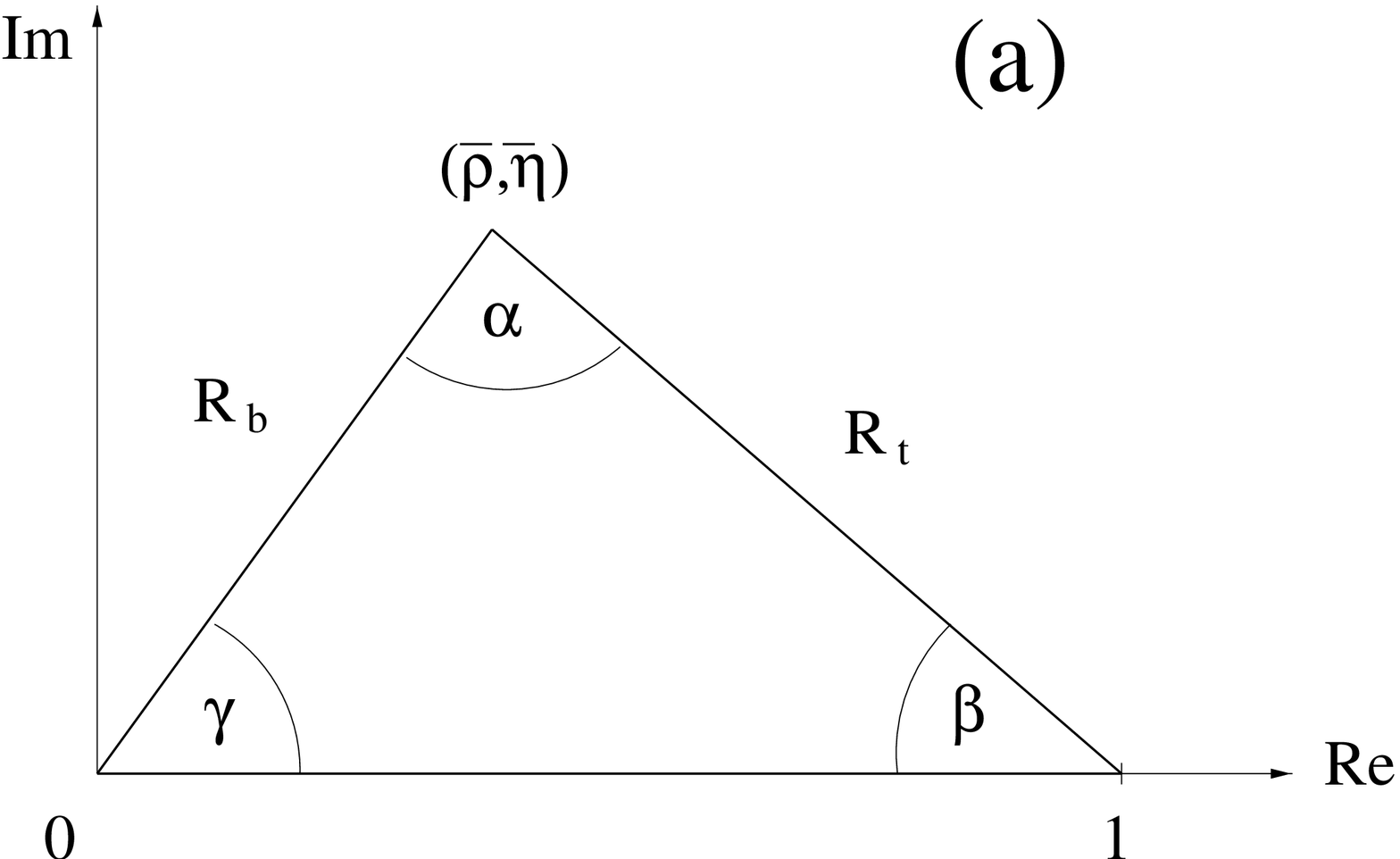,height=2.0in}
\psfig{figure=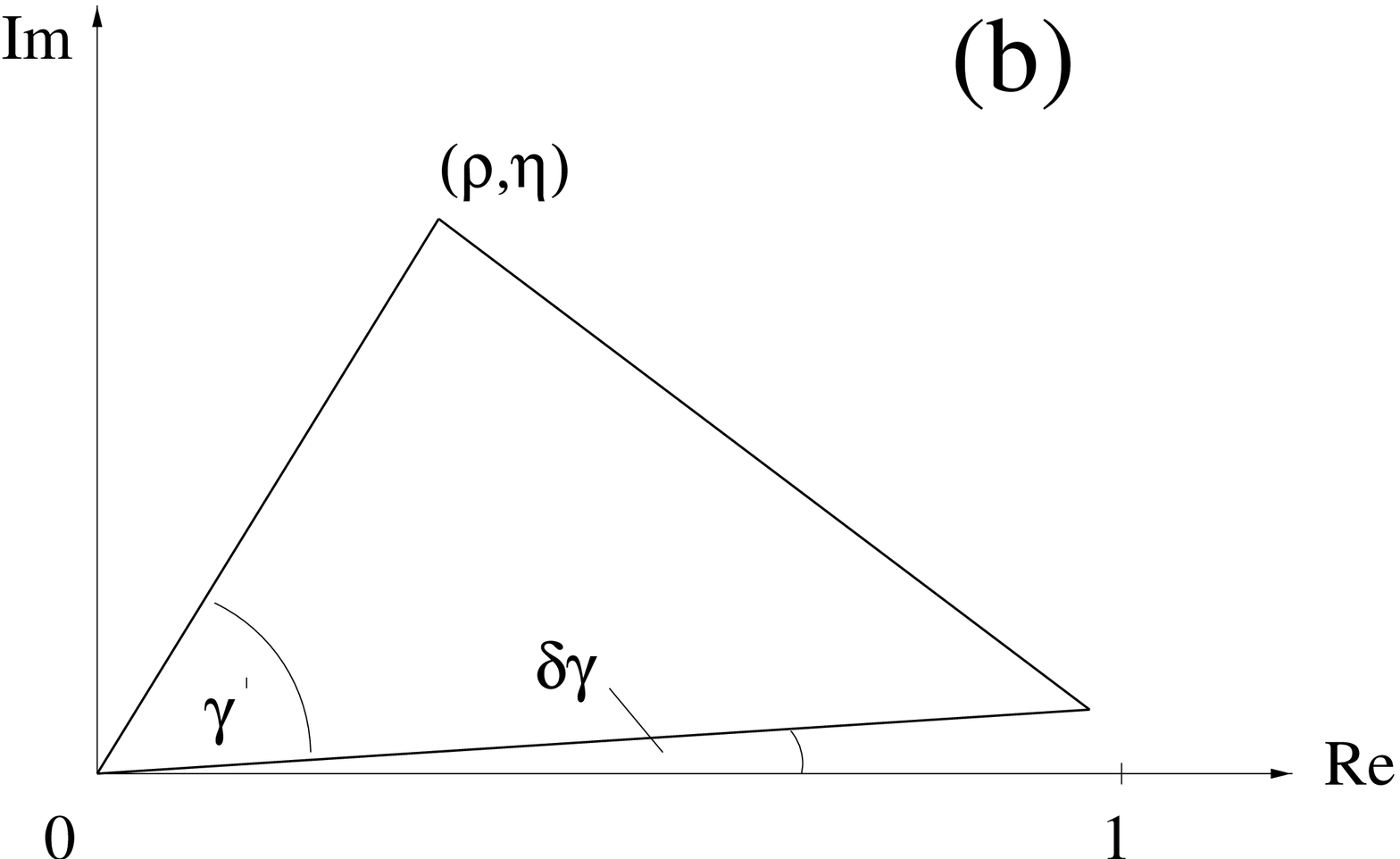,height=2.0in}}
\caption{The unitary triangles $(db)$ and $(ut)$ in complex space.
\label{fig:triangle}}
\end{figure}

The unitarity of the CKM matrix implies that there are six orthogonality
\cite {aleksan} 
conditions between any pair of columns or any pair of rows of the matrix:
\begin{eqnarray*}
\begin{array}{cccr}
V_{ud}V^*_{ub}+V_{cd}V^*_{cb}+V_{td}V^*_{tb}&=&0  &(db) \\
V_{us}V^*_{ub}+V_{cs}V^*_{cb}+V_{ts}V^*_{tb}&=&0  &(sb) \\
V_{ud}V^*_{us}+V_{cd}V^*_{cs}+V_{td}V^*_{ts}&=&0  &(ds) \\
V_{ud}V^*_{td}+V_{us}V^*_{ts}+V_{ub}V^*_{tb}&=&0  &(ut) \\
V_{cd}V^*_{td}+V_{cs}V^*_{ts}+V_{cb}V^*_{tb}&=&0  &(ct) \\
V_{ud}V^*_{cd}+V_{us}V^*_{cs}+V_{ub}V^*_{cb}&=&0  &(uc)
\end{array}
\end{eqnarray*}
where the quark pair in parenthesis indicates the row or column
under consideration.  Each of the orthogonality conditions requires 
the sum of three complex numbers to vanish and can be represented 
as a so called "unitarity triangle" in complex space. Only triangles
$(db)$ and $(ut)$ have three large angles.  The two triangles are
drawn in Fig. \ref{fig:triangle}.  

The apex of the triangles
have coordinates ($\rho$, $\eta$) or
($\bar{\rho}, \bar{\eta}$)
as shown in figure 1.  The three angles of the unitarity
unitarity triangle $(db)$ are :
$$
\alpha = \arg \left[- \frac {V_{td}V^*_{tb}}{ V_{ud}V^*_{ub}}\right] 
~~~~~~\beta = \arg \left[- \frac {V_{cd}V^*_{cb}}{ V_{td}V^*_{tb}}\right] 
~~~~~~\gamma = \arg \left[- \frac {V_{ud}V^*_{ub}}{ V_{cd}V^*_{cb}}\right]
$$
while the three angles of the triangle $(ut)$ are:
$$
\alpha^{\prime} = \arg \left[- \frac {V_{td}V^*_{ud}}{ V_{tb}V^*_{ub}}\right] 
~~~~~~\beta^{\prime} = \arg \left[- \frac {V_{ts}V^*_{us}}{ V_{td}V^*_{ud}}\right] 
~~~~~~\gamma^{\prime} = \arg \left[- \frac {V_{tb}V^*_{ub}}{ V_{ts}V^*_{us}}\right]
$$

Measurements of weak decays of $B$ hadrons determine the magnitudes of 
the sides of the triangles:
$$
R_b=\left( 1-\frac{\lambda^2}{2}\right) \frac{1}{\lambda} 
\left| \frac{V_{ub}}{V_{cb}} \right|
~~~~~~~~~~~
R_t=\frac{1}{\lambda} 
\left| \frac{V_{td}}{V_{cb}} \right|
$$
while CP asymmetries and rates of $B$ meson decays determine 
the three angles.  

The angles of the two unitarity triangles are also related by
the following equations:
$$
\alpha^{\prime}=\alpha
~~~~~~~\beta^{\prime}=\beta+\arg{[V_{ts}]}
~~~~~~~\gamma^{\prime}=\gamma-\arg{[V_{ts}]}
$$
Therefore if we define $\delta \gamma=\gamma^{\prime}-\gamma$
then $\delta \gamma=\eta \lambda^2$.
Therefore the two triangles are identical 
at leading order in the Wolfenstein expansion. However we expect that future
high statistics dedicated experiments will be sensitive to these differences
and further probe the CKM matrix.

At hadron colliders the $B_s$ mesons is copiously produced
and therefore there is an opportunity to measure the $B_s$ asymmetries. The 
unitarity triangle $(s,b)$ is squashed since the first side is much shorter 
than the other two and the opposing angle: 
$$
\beta_s=\arg \left[- \frac {V_{ts}V^*_{tb}}{ V_{cs}V^*_{cb}}\right]= \lambda^2 \eta +
\mathcal{O}(\lambda^4)  
$$
is of the order of one degree.
Therefore in the standard model we expect the $B_s$ CP
asymmetries to be smaller than in the $B$ system.
On the other hand these asymmetries are sensitive to new physics
and therefore the measurement of $\beta_s$ is especially important.

The goal of B physics in the next decade is to overdetermine the CKM
matrix to test the consistency of the standard model and hopefully discover
new physics.
Signs of physics beyond the standard model could appear as:
\begin {itemize}
\item The standard model predictions for the branching fractions and
the CP asymmetries will disagree with the experimental measurements
and $ \alpha + \beta + \gamma \neq \pi$
\item
The values of $\rho$ and $\eta$ determined from $B$ decays disagree
with the values obtained from $K$ decays
\item Decays forbidden or rare in the standard model occur at larger rates
than expected.
\item CP asymmetries larger than expected in the standard model. 
\end{itemize}
$B_s$ mesons are expected to have a special role in this effort
since it has been suggested that $\beta_s$ can be measured 
in $B_s\rightarrow J/\psi \eta (^\prime)$.

\section{$B$ production at hadron colliders}

Electron positron \bm-factories running at the $\Upsilon$(4S) which is at
the threshold for open beauty production have been
very effective in the study of $B$ mesons. Recently with the advent of 
precise silicon detectors crucial measurements have also been performed
at LEP and at hadron colliders. 

B meson production at the $\Upsilon$(4S) is dominated by $B^0\bar{B}^0$ and 
$B^+B^-$ final states. The cross section for $b\bar{b}$ production
$\sigma_{b\bar{b}}$ is 1.15 nb and the signal to background defined 
as the ratio $S/B=\sigma_{b\bar{b}}/\sigma_{total}$ 
is 0.25. The \bm-factories operate
in an asymmetric beam configuration that allows for the production 
of the $\Upsilon$(4S) boosted along the beam axis. The separation between 
the two $B$'s along the beam axis is 200 and 260 $\mu m$ at Belle and 
BaBar respectively.
These machines are ideally suited to study CP violation in 
$J/\psi K^0_s$, for the precision determination of $V_{cb}$ and $V_{ub}$ in 
semileptonic decays and rare decays of B mesons into final states
which include neutral pions and photons such $B\rightarrow K^* \gamma$.  

Electron positron colliders running at the $Z^0$ such as LEP and
SLD have the advantage 
of a larger boost that allows a better separation 
of the decay vertices of charm 
and beauty hadrons. The B meson mean decay length at 
the $Z^0$ is 2.7 mm. The cross section 
is $\sigma_{b\bar{b}}$ of 7 nb and the S/B is similar to the one
obtained by the \bm-factories. At LEP and SLD, because of the higher 
energy available, all B hadron species are produced including 
$B_s$, $B_c$, $\Lambda_b$ and $\Omega_b$. These colliders have contributed 
significantly 
to our current knowledge of $B$ hadrons lifetimes and mixing.

The cross section for $b\bar{b}$ production at the Tevatron, a hadronic machine
running at a center of mass energy of 2.0 TeV, is about 100 $\mu b$. Unfortunately the
background is also large and the S/B is only about $0.2$\%. 
At a center of mass energy of 14 TeV, at which 
the Large Hadron Collider will operate, the $b\bar{b}$ cross section and the S/B will
increase to 400 $\mu$b and $0.4$\%. Specialized triggers are necessary in order
to write the events of interest to tape. The average decay length of 
produced $b$ hadrons is a few mm because of the 
momentum boost. Similarly to LEP, a full spectrum of $b$ mesons
and baryons are produced at hadronic machines.

\begin{figure}
%\rule{5cm}{0.2mm}\hfill\rule{5cm}{0.2mm}
%\vskip 2.5cm
%\rule{5cm}{0.2mm}\hfill\rule{5cm}{0.2mm}
\centerline{\psfig{figure=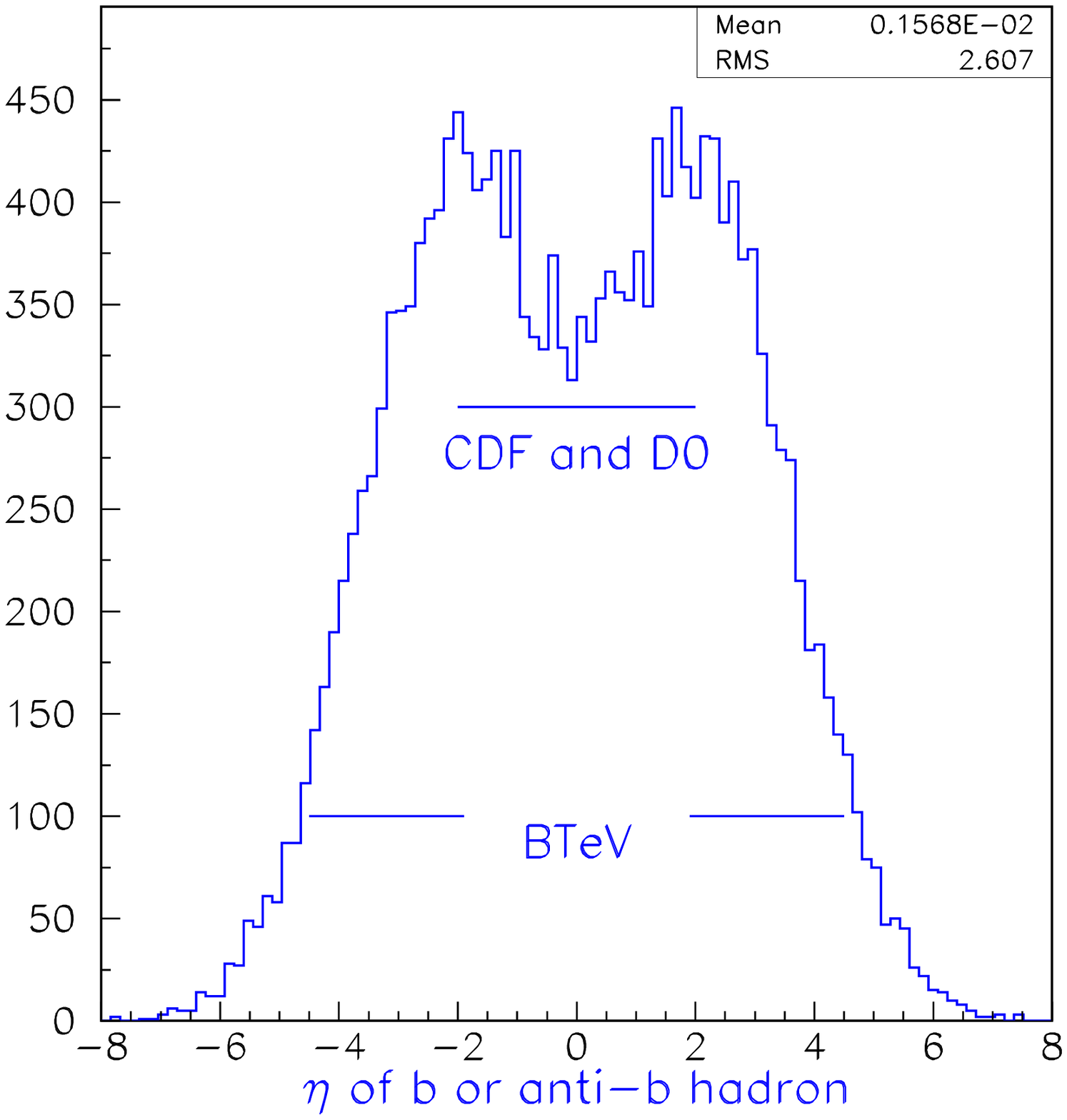,height=2.5in}
\psfig{figure=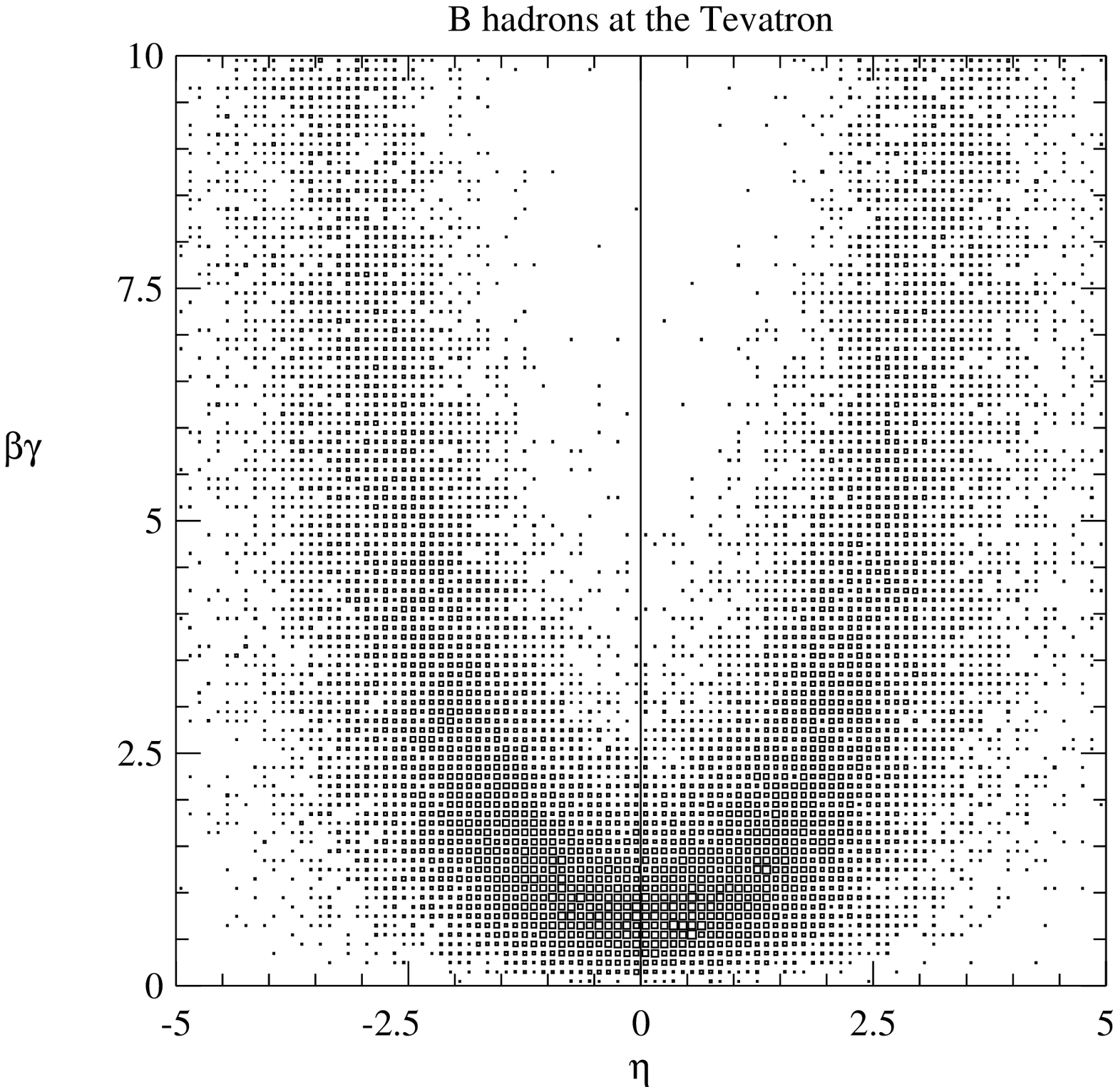,height=2.5in}}
\caption{Left: The $B$ production cross section versus $\eta=-\ln(\tan\theta/2)$.
Right: The $B$ production cross section as a function of the lorentz boost 
$\beta \gamma$ and $\eta$.
\label{fig:prod}}
\end{figure}

Two kinds of $b$ experimental configuration 
are considered for hadron colliders as shown in Fig. \ref{fig:prod}. 
The first is a central 
all-purpose detector such as CDF and D0.
This configuration is also implemented in CMS and ATLAS which will take 
data at the LHC. The bulk of the $b\bar{b}$ production is concentrated in the
central rapidity region. Nonetheless forward going $B$ mesons have much higher
momentum and therefore longer decay 
length than $B$ mesons produced in the central region. 
Dedicated forward $b$-experiments
such as B-TeV and LHC-b are in the planning stage or under
construction for operation at the Tevatron and the LHC respectively.
Forward experiments exploit the correlation between 
the direction of the produced $b\bar{b}$ and the boost direction. Moreover
dedicated experiments can provide better particle identification
and neutral pion identification.

\section{Experimental issues}

\subsection{Trigger}
The primary experimental issue of $B$ physics at hadron collider is the
triggering. The $b\bar {b}$ production rate at $B$ factory operating
at a luminosity $L=10^{34}cm^{-2}s^{-1}$ is about 10 Hz while it is 
about 20 KHz at the Tevatron and 100 KHz 
at the LHC at $L=10^{33}cm^{-2}s^{-1}$.
During Run I of the Tevatron the CDF and D0 triggers relied on the presence of 
single and dilepton triggers
such as $B\rightarrow J/\psi X$ and then $J/\psi \rightarrow \mu^+ \mu^-$. 
Although the branching fraction for 
$B\rightarrow J/\psi X$ is only 1.15 \% hadronic experiments have an advantage 
with respect to the \bm-factories because of the large production cross 
section available at hadronic machines. 

In Run II of the Tevatron, that started in spring 2001, CDF has also 
implemented a new high 
precision trigger that allows for the selection of events with detached vertices. 
These triggers based on precise vertex information
from silicon detectors will also be adopted by CMS,
ATLAS, LHCb and BTeV. They allow a factor of about 100 enrichment in $b$ content 
at the trigger level. Moreover this displaced vertex trigger
method is crucial for selecting hadronic decays
such as $B\rightarrow  \pi^+ \pi^-$ and 
$B_s\rightarrow D^-_s \pi^+$.

\subsection{Flavor Tagging}

The measurement of CP violation and mixing requires the identification
of the flavor of the 
B meson at the time of production. Tagging algorithms are evaluated 
in terms of efficiency $\epsilon$ for determining the flavor tag ($b$ or $\bar{b}$)
and the probability that the tag is correct.  The quality of the tag is evaluated
by defining the tagging dilution as $D=(N_R-N_W)/(N_R+N_W)$ where $N_R(N_W)$ 
is the number of right (wrong) tags. The observed asymmetry is reduced
with respect to the true asymmetry by the dilution $D$ and 
$A^{obs}_{CP}=DA_{CP}.$  Therefore the maximum sensitivity can be achieved 
when the dilution is large and a perfect tagging algorithm will have
a dilution $D=1$.
The statistical uncertainty on $\sin 2\beta$ is inversely proportional to 
$\sqrt{\epsilon D^2}$. The statistical uncertainty is also proportional to
$\sqrt{ S^2/(S+B)}$ where $S$ is the number of signal 
events and $B$ is the number of backgrounds events. Therefore the data 
sample should be chosen to maximize the signal and minimize the background.

Several tagging methods are used to improve $\epsilon D^2$. 
The opposite side tagging algorithms identify the flavor of the
{\it opposite}
B in the event at the time of production. If one knows that
the other $b$ hadron contains a $b$ quark then the signal $B$
meson must contain at $\bar{b}$ quark. Several methods
of opposite side tagging can be employed: soft lepton tag (SLT), soft kaon tagging 
and jet-charge tag (JETQ).

The soft lepton tag associates the charge of the lepton
($e$ or $\mu$) from semileptonic decays with the flavor of the parent B
meson as $b\rightarrow X \ell^-$ compared 
to $\bar{b} \rightarrow X \ell^+$. Since we are tagging the opposite
$B$ meson, its flavor is anti-correlated with the flavor of the $B$-meson that
decays to the mode under study. Hence a $\ell^-$($\ell^+$) tags a $\bar{b}$($b$)
like $B^0$($\bar{B^0}$). The branching fraction for semileptonic decays
is about 10\% into each $e$ and $\mu$ channels. There is also dilution
because of sequential decays where a $b$ hadron decays into a $c$ hadron which then
decays semileptonically. However the leptons from direct 
and sequential decays
have different kinematic properties and a good separation can be achieved.
Further dilution is caused by mixing since 17.4\% of the $B^0$ will oscillate to 
a $\bar{B}^0$ before decaying. Moreover the $B_s$ is fully mixed and will not provide
any tagging power. 

"Jet charge" or JETQ, tags the
$b$ flavor by measuring the average charge of the opposite
side jet which is calculated as:
$$
Q_{jet}=\frac{\sum_i q_i (\vec{p_i} \cdot \hat{a})}{\sum_i \vec{p_i}\cdot \hat{a}}
$$
where $q_i$ and $\vec{p_i}$ are the charge and momentum
of track $i$ in the jet and $\hat{a}$ is the unit vector along the 
jet axis. On average the sign of the jet charge gives the 
flavor of the $b$ quark that produced the jet.

Kaon tagging exploits charge of the kaon in the away side because
of the decay chain $b \rightarrow c \rightarrow X K^-$ compared to
$\bar b \rightarrow \bar c \rightarrow X K^+$. Since the product
of branching fractions is large this tagger has a larger efficiency than
the SLT but lower dilution. Excellent particle identification
is necessary.

The same side tagging method or SST relies on the correlation
between the $B$ flavor and the charge of the nearest pion in the fragmentation
chain. Such a correlation can arise from the fragmentation processes which form a B
meson from a $\bar b$ quark and from
the decay of an excited B meson state ($B^{**}$).
In the fragmentation a $\bar b$ quark forming a $ B^0$ can combine
with a $d$ in the hadronization leaving a $\bar d$ which can form a $\pi^+$
with a $u$ quark from the sea. The excited B state
will decay $B^{**+} \rightarrow B^{(*)0}\pi^+$.
Therefore in both cases a $B^0$ ($\bar B^0$) meson is associated with a
positive (negative) particle respectively.

The dilution parameters for all tagging algorithms can be measured
on calibration samples. At hadronic machines the strong interaction 
creates $b \bar b$ pairs
at sufficiently high energy that the B mesons are largely
uncorrelated. For example, the
$b$ quark could hadronize as a $\bar{B}^0$ while the
$\bar b$ could hadronize as a $ B^+$, $B^0$, or $B^0_s$ meson. Therefore
we can use a samples of $B^{\pm} \rightarrow J/\psi K^{\pm} $
decays to measure the tagging dilutions for the opposite side algorithms.
The performance of the same side tagging methods is usually evaluated by tagging
$B \rightarrow \nu \ell D^{(*)}$ decays and by measuring
the time dependence of $B^0 \bar{B}^0$ oscillations in this high statistics
sample and in a lower statistics sample of
$B \rightarrow J/\psi K^{*0}$. The tagging methods have different 
characteristics. The lepton tagging has 
relatively low efficiency but good dilution. The same side tagging and the jet charge 
tagging are more efficient but have lower dilution. Kaon tagging performance will
depend crucially on the particle identification capabilities of the specific detectors.
The different tagging information can be combined to obtain 
a more powerful performance. Such combination must account for 
correlations. In general flavor tagging at 
hadron collider experiment achieves an effective efficiency 
$\epsilon_{eff}=\epsilon D^2 \approx 10$ \% while at a b-factory
it is about 30\%. These expectations are confirmed by the 
CDF tagging performance in run I where CDF used most of these 
methods but the kaon tagging for the measurement of $B_d$
mixing and $\sin 2\beta$.

\subsection{Measurement Error on Proper Decay Time}

The $b$ hadron proper decay time is evaluated by measuring the 
decay length which is the distance between the primary
vertex and the secondary vertex where the $b$ hadron decayed. The proper decay
time is $t=Lm/pc=L/\beta \gamma c$ where $L$ is the decay 
length, $p$ and $m$ is the reconstructed momentum and mass of the 
$b$ hadron, $c$ is the speed of light, and $\beta$ and $\gamma$ 
are the $b$ hadron Lorentz parameters. The
uncertainty in the primary vertex position, the secondary vertex position, and the $b$
hadron momentum all contribute to the measurement error on the decay length.
Since the secondary vertex position uncertainty is larger 
than the primary vertex and much larger than the momentum uncertainty
then:

$$
\sigma_{t}\approx \sigma^{secondary}_L/(\beta\gamma c)
$$

The proper time resolution, $\sigma_{t}$, of the forward 
detectors in hadronic machines such as BTeV and LHCb  
is 40 $f{\rm s}$ while it
is about 900 $f{\rm s}$ for the \bm-factories.

\subsection{Performance of the displaced vertex trigger at CDF}

The CDF silicon vertex trigger (SVT) has opened a new era 
of B-physics opportunities at a hadron collider machine.
It consists of a real-time tracker capable of reconstructing 2D silicon 
tracks with the offline quality resolution that is an essential
tool to discriminate $b \bar {b} $ and $c \bar{c}$ from 
hadronic background. The results obtained up to now are benchmarking the 
capability of the trigger but they are not yet fully representative of the
physics that this triggers will make possible.

The CDF trigger has a three level architecture. At Level 1
the decision taken within 5.5 $\mu$s is based on objects
such as electrons, muons, jets and tracks.  The tracks are reconstructed
by the extremely Fast Tracker (XFT), an asynchronous, parallel and pipelined
track processor that reconstructs the tracks on the plane 
transverse to the beam by using the information of the open cell
drift chamber (COT). At Level 2 the SVT takes the list of tracks 
found by the XFT 
and adds to this the information from four layers of the 
silicon detector.  The results of the 
association of the silicon hits to each XFT tracks yield
the track parameters ($d_0$, $\phi$, $p_t$) with almost offline resolution. The 
impact parameter $d_0$ of the track 
with respect to the beam spot is then used to select events with 
displaced vertices.
During the beginning of run II CDF has used the level-1 and level-2 
track processors to select events that have two tracks 
with impact parameter larger than 100 $\mu$m and 
decay length in the transverse plane greater than 200 $\mu$m.
The trigger has been performing well. Currently 
the Run II luminosity has been about $10^{31}$ cm$^{-2}$ s$^{-1}$ 
and the Level-1 rate is $\approx 3$ KHz and the level 2 rate 
is 50 Hz.  When the luminosity will increase further cuts will have 
to be introduced to keep the rate at an acceptable level.

\begin{figure}
%\rule{5cm}{0.2mm}\hfill\rule{5cm}{0.2mm}
%\vskip 2.5cm
%\rule{5cm}{0.2mm}\hfill\rule{5cm}{0.2mm}
\centerline{\psfig{figure=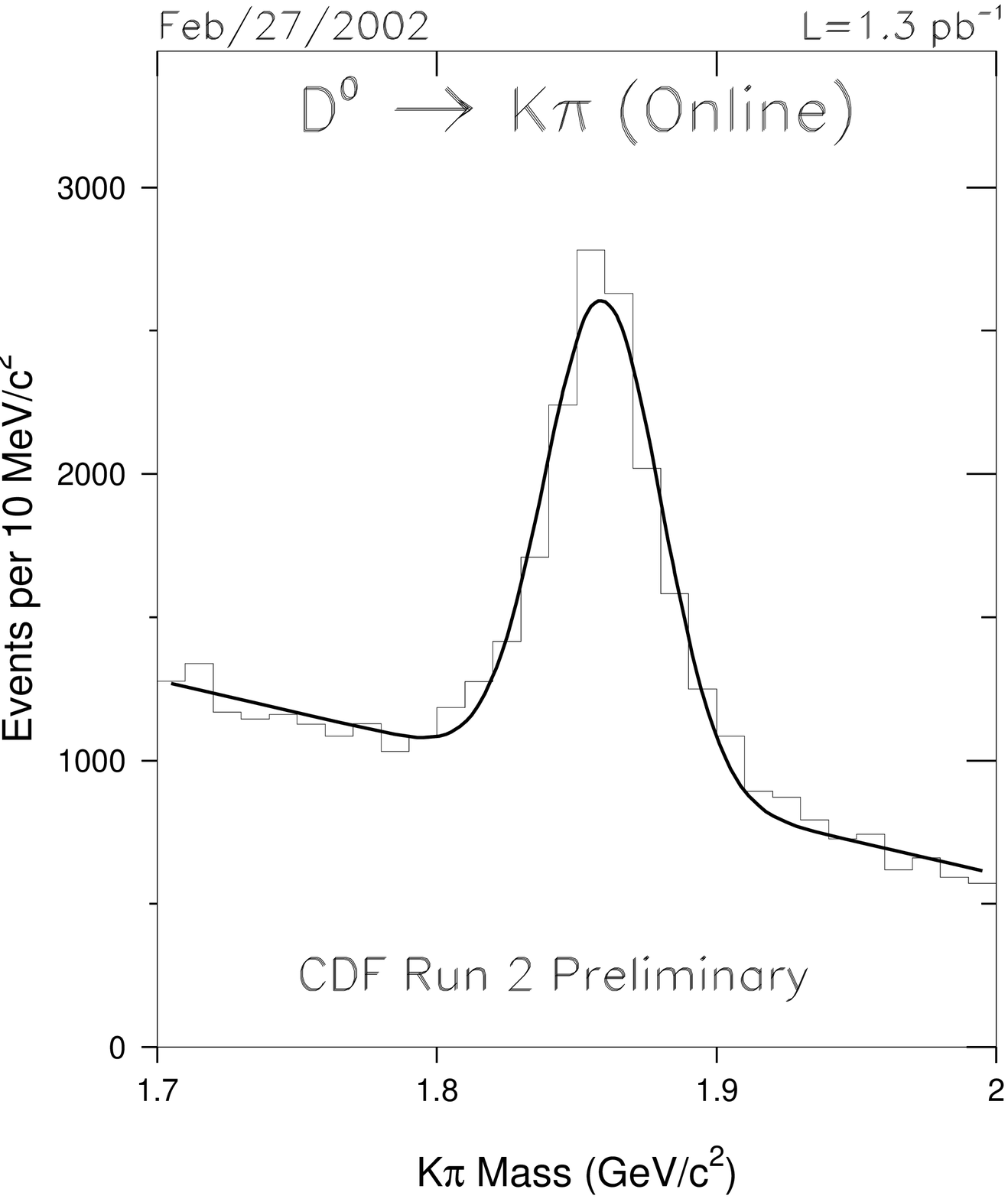,height=2.5in}
\psfig{figure=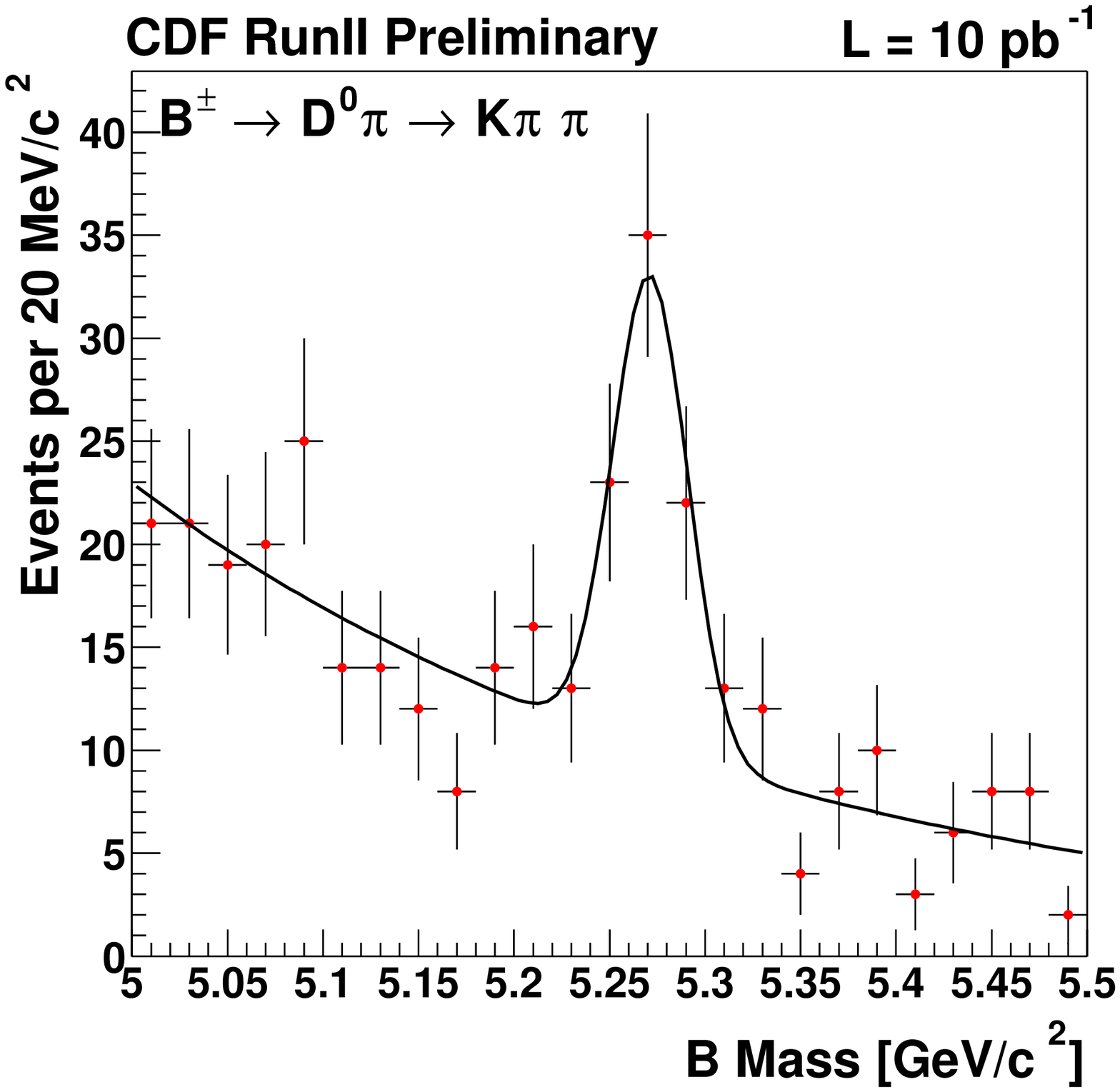,height=2.5in}}
\caption{Left: The $D^0 \rightarrow K \pi$ invariant mass distribution
measured online at CDF. Right: $B \rightarrow D^0 \pi$ invariant mass 
distribution in decays collected with the SVT 
trigger.
\label{fig:hadronic}}
\end{figure}

The cross section for charm is large and CDF can monitor 
the SVT performance by reconstructing $D^0\rightarrow K^- \pi^+$ online.
The trigger has also allowed CDF to collect $B\rightarrow h^+ h^-$ where
$h$ is a kaon or pion candidate and 
$B\rightarrow D^0 \pi$ events.
The online $D^0$ mass peak and the invariant mass distribution 
of $B \rightarrow D^0 \pi$ candidates 
are shown in fig. \ref{fig:hadronic}.

\section{Crucial measurements}

In order to evaluate the reach of hadronic machines
a series of specific measurements such as
$\sin 2\beta$, measurement of $B_s$ mixing and
the determination of $\alpha$ and $\gamma$ will be discussed.

\subsection{$B_s$ mixing}

The $B_s$ is unique to hadron colliders and therefore
one of the physics goals of the Tevatron is to measure in detail
the $B_s$ mixing amplitudes including 
$x_s$, $\Delta \Gamma_s$ and the $B_s$ mixing phase $\phi_s=-2\beta_s$. 

These measurements are crucial
not only in terms of checking the CKM unitary triangle but also
because they are expected to be sensitive to physics beyond the 
standard model. 

The measurement of the ratio of the $B_s$ and $B_d$
mixing frequencies $x_s$ and $x_d$ would benefit enormously the 
determination of the right side of the $bd$ unitarity 
triangle $R_t$ since:
$$
\frac{\Delta m_s}{\Delta m_d}= 
\frac{m_{B_s}}{m_{B_d}} \frac{|V_{ts}|^2}{|V_{td}|^2} 
\frac{F^2_{B_s}B_{B_s}}{F^2_{B_d}B_{B_d}}
$$ 
In fact the measurement of $x_s$ will constraint the
right side of the unitary triangle to better than 6\% 
since the uncertainties on the hadronic
matrix elements partially cancel in the
ratio $\xi=F_{B_s}\sqrt{B_{B_s}} / F_{B_d}\sqrt{B_{B_d}}$. The world average
which is dominated by the LEP and SLD constrains $\Delta m_s$ 
to be $>19$ at 95 \% confidence level (CL).
Analyses of the unitary triangle\cite{parodi} predict
$22 < x_s < 30.8$. To measure such a rapid oscillation
frequency, excellent decay time resolution is 
crucial. 
%The expected sensitivity to $\Delta m_s=x_s/\tau_{B_s$ is given by :
%$$
%{\rm Sig}(\Delta m_s)=\sqrt{\frac{N \epsilon D^2}{2}}
%\exp^{-(\Delta m_s \sigma_t/ct)^2/2} \sqrt{\frac{S}{S+B}} 
%$$
%Where N=S is the number of reconstructed $B_s$ signal, S/B is the 
%signal to background, $\epsilon D^2$ is the effective
%flavor tagging efficiency and $\sigma_t$ is the proper time 
%resolution.

Semileptonic decays such as $B_s \rightarrow D_s \ell \nu$ 
suffer from a large smearing of the decay time 
resolution due to the the missing neutrino momentum 
and can only probe up to $x_s=30$ with 2 fb$^{-1}$. 
Better performance can be obtained by studying 
hadronic decays such as $B^0 \rightarrow D^-_s \pi^+$
that can be triggered upon using secondary vertex triggers.
The current CDF II silicon detector is expected to achieve
at least a proper time resolution of 60 fs. The $ct$ resolution
could improve to 45 fs if the
low mass silicon layer closer to the beam pipe (L00)
can be effectively integrated.
Currently this layer is not used in the tracking.
The flavor tagging effective efficiency in $B_s$ decays
is expected to be about $11.3$ \% at CDF in Run II.   

\begin{figure}
%\rule{5cm}{0.2mm}\hfill\rule{5cm}{0.2mm}
%\vskip 2.5cm
%\rule{5cm}{0.2mm}\hfill\rule{5cm}{0.2mm}
\centerline{\psfig{figure=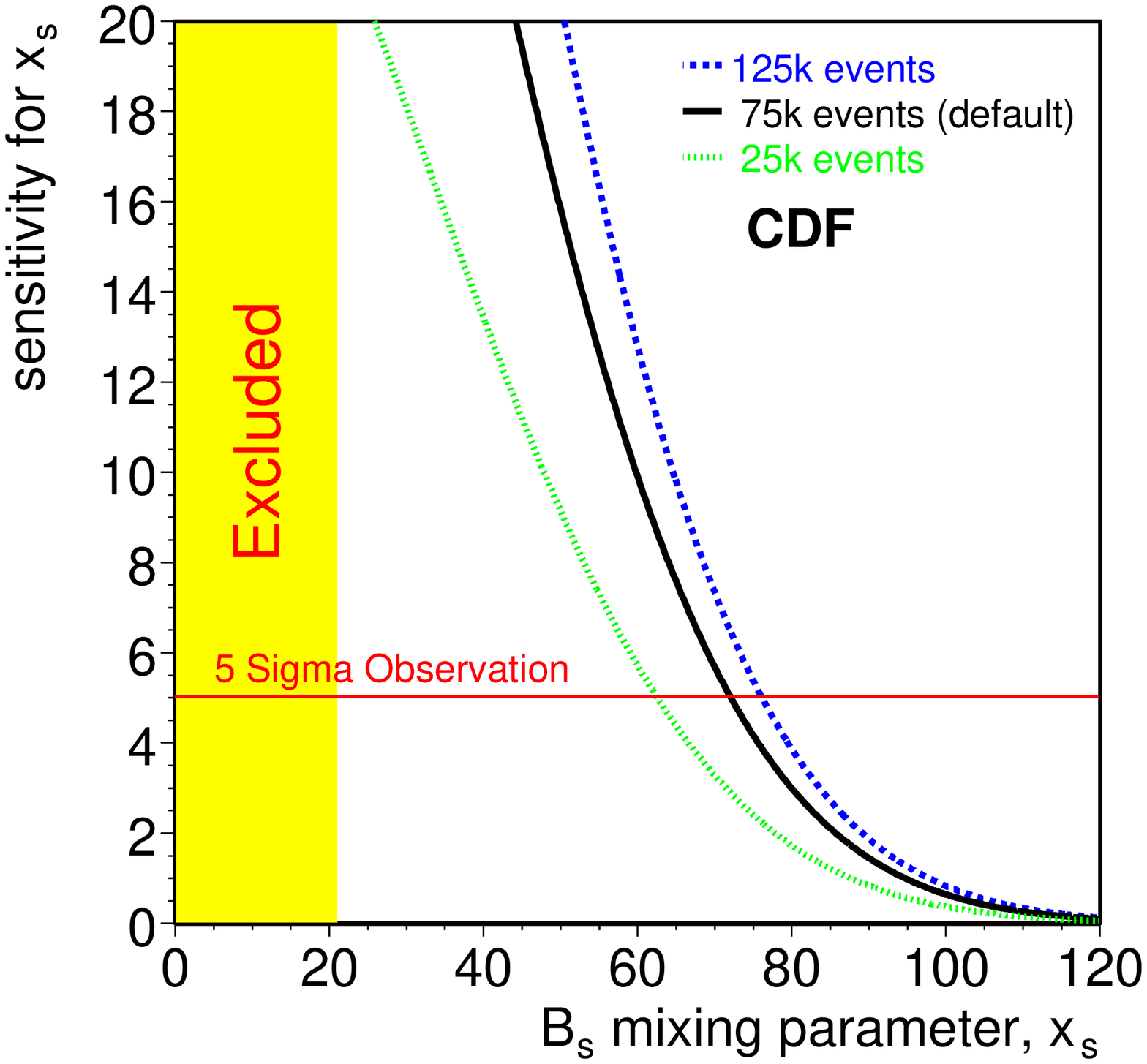,height=2.5in}
\psfig{figure=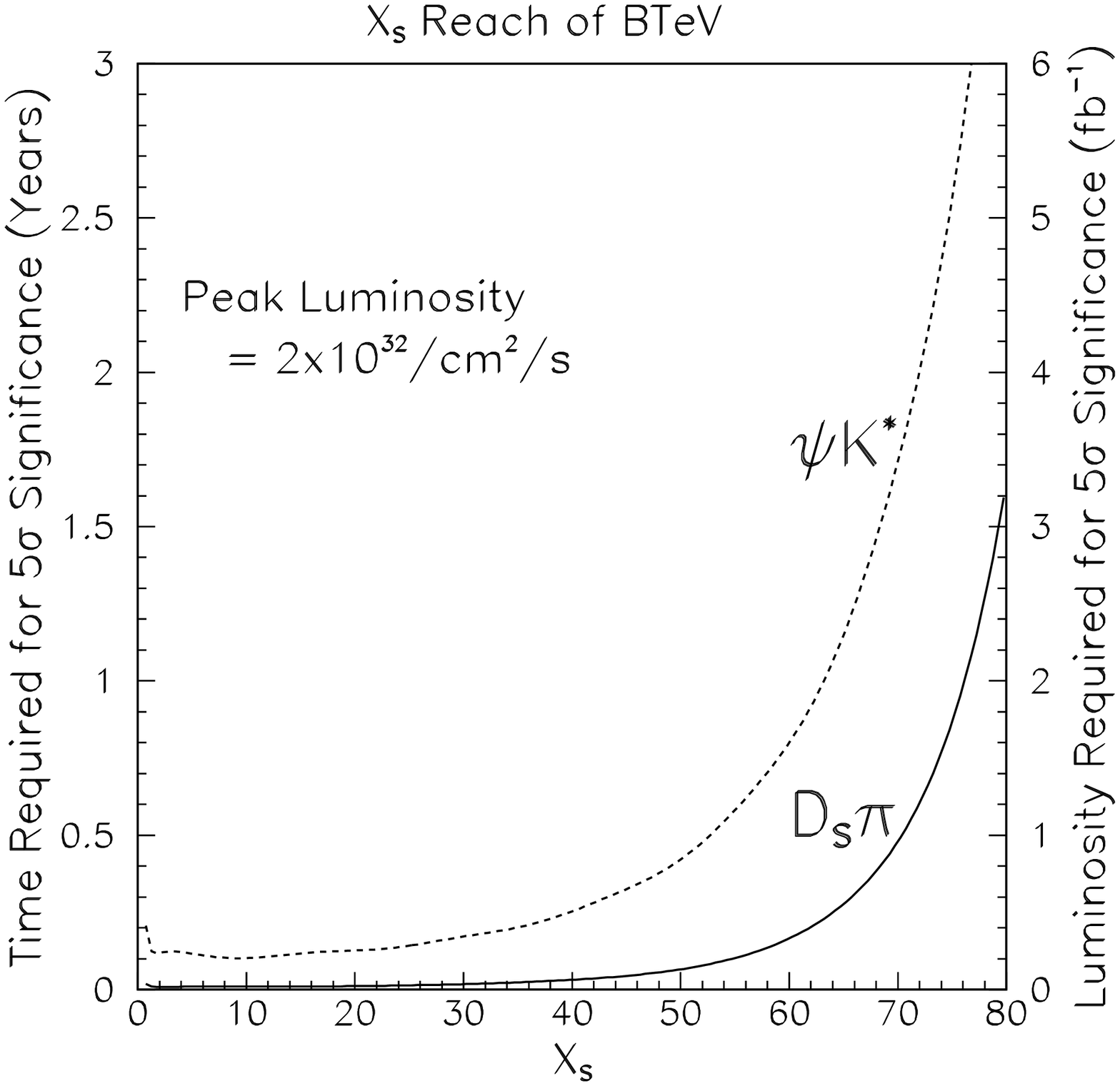,height=2.5in}}
\caption{ Left: The $x_s$ reach of BTeV
for different assumptions of the 
even yield. The curve assumes
a $S/B=1$ and $ct$ resolution of 60 fs.
Right: The $x_s$ reach of the BTeV detector. The curve 
indicates the time needed to make a 5$\sigma$ measurement.
\label{fig:bmix}}
\end{figure}

For $x_s$ the expected
reach depends strongly on 
the number of $B_s$ events that can be reconstructed in 
the $B_s \rightarrow D^-_s \pi^+ ( \pi^+ \pi^- ) $ samples,
the $S/B$, the $ct$ resolution, the effective $b$ tagging and 
the SVT efficiency. In the most optimistic 
scenarios with $N(B_s)= 75,000$, 
$\epsilon D^2$=11.3 \% and S/B=2
CDF expects to probe up to $x_s$ of about 74
with 2 $fb^{-1}$ of data. This expectation decreases to 
$x_s$ of about 50 in a more pessimistic scenario. 
The CDF reach with L00 is shown in figure \ref{fig:bmix}.

The statistical error on $x_s$ is related
to the statistical error on the observed mixing amplitude,
therefore once a measurement of $x_s$ is obtained at the 
5 $\sigma $ level the 
$\sigma (x_s)=\frac{1}{5\sqrt{2}}$ and $\sigma (x_s)/x_s<1$\%.

BTeV and LHCb will obtain a similar sensitivity to $x_s$.
For example BTeV expects to observe any value of $x_s$ less than 75 in one year 
of running which is equivalent to 2 fb$^{-1}$. Therefore we expect that once
$x_s$ is observed then it will 
be measured very easily and precisely.

\subsection{The determination of $\sin 2\beta$}

The angle $\beta$ can be determined by studying many different
types of b decay modes.
The best determination of $\sin 2\beta$ can be achieved by studying color
suppressed decays such as $b \rightarrow c \bar c s$.
The golden mode is the decay $B^0/\bar B^0 \rightarrow J/\psi K^0_s$ since
the dominant penguin contribution has the same weak phase as the tree
amplitude\cite {bluebook}. The only term with a different weak phase
is suppressed by ${\cal O}(\lambda^2)$. Therefore the extraction of
$\beta$ from the measurement of the asymmetry in
$B^0/\bar B^0 \rightarrow J/\psi K^0_s$ suffers negligible
theoretical uncertainties.

At hadron colliders the golden mode
$B \rightarrow J/\psi K^0_s$ is especially interesting experimentally
because the $J/\psi \rightarrow \mu^+ \mu^-$ decay mode gives a
unique signature and allows for a powerful trigger.
Therefore the decay $B^0 \rightarrow J/\psi K^0_S$ can be
reconstructed with an excellent signal to background ratio. The
interference of the direct decays $B^0 \rightarrow J/\psi K^0_S$ and
$B^0 \rightarrow \bar B^0 \rightarrow J/\psi K^0_S$ gives rise to a CP
asymmetry that measures $\sin 2\beta$ :
$$
A_{CP}(t)=\frac{ \bar B^0(t) - B^0(t) }{ \bar B^0(t) + B^0(t) }
=\sin 2\beta \sin \Delta m_d t
$$
This formula neglects the possibility of any direct CP violation 
which will introduce a term dependent on $\cos \Delta m_d t$. The time 
integrated asymmetry is:
$$
A_{CP}=\frac{x_d}{1+x_d^2}\sin2\beta
$$
To measure the asymmetry we have to identify the flavor of the
$B$ meson at the time of production. The statistical 
uncertainty on $\sin 2\beta$ is inversely proportional to
$\sqrt{\epsilon D^2}$ and to
$\sqrt{ S^2/(S+B)}$ where $S$ is the number of signal
events and $B$ is the number of background events within three
standard deviations of the B mass. Therefore the data sample should be
chosen to maximize the signal and minimize the background.
CDF measured $\sin 2 \beta = 0.79 ^{+0.41}_{-0.44}$ (stat + sys)
using the run I data sample \cite{sin2b}. Recently the Run 1 data sample
has been re-analyzed to improve the tagging performance. The latest 
CDF value of $\sin 2 \beta $ is $0.91 ^{+0.37}_{-0.36}$ (stat + sys).

In run II CDF expects to improve the effective tagging efficiency 
from $\epsilon_{eff}=(6.3 \pm 1.7 )$ \% to $\approx 9.1$ \%. The 
number of reconstructed
$B\rightarrow J/\psi K^0_s$ is expected to increase because of the increase
in the center of mass energy of the Tevatron from 1.8 to 1.96 GeV/c$^2$,
increased muon coverage, and lower di-muon momentum threshold in  the
trigger. Monte Carlo studies estimated an increase of a factor of 50 
in the $J/\psi K^0_s$ yield over the 400 events found in run 1A. We expect 
this improvement to result in an conservative 
error $\sigma(\sin 2 \beta) \approx 0.05$.

The D0 collaboration has also studied their run II capabilities
for the measurement of $\sin 2 \beta$. The result of their 
Monte Carlo studies shows an expected data sample of about 
34,000 $B \rightarrow J/\psi K_s$ events in 2 fb$^{-1}$. They 
expect an effective tagging efficiency of about 10\% and 
an error $\sigma(\sin 2 \beta) \approx 0.04$.

BTeV at the Tevatron will reconstruct about
80,500 $B\rightarrow J/\psi K^0_s$ after running
for one year at a luminosity of $2\times 10^{32}$ cm$^{-2}$ s$^{-1}$.
This yields an error on $\sin 2\beta$ of about 0.025.

The measurement of $\sin 2 \beta$ will continue to be improved by 
ATLAS and CMS. The mode $B\rightarrow J/\psi K^0_s$
was considered as one of the LHC benchmark modes in a recent study \cite{LHC}.   
Combining the 
statistical samples after 
3 years of data taking by ATLAS and CMS with 5 years 
of running at LHCb we expect a statistical error on $\sin 2 \beta$
of 0.005. This precision is an order of magnitude better than the 
expected statistical precision that the $e^+e^-$ \bm-factory are 
expected to achieve by 2005 when they will have collected 0.5 ab$^{-1}$.
The large data samples expected at the
LHC will allow a probe for a direct CP violating phase.
Fitting the data with an additional term degrades the precision
on $\sin 2 \beta$ by $\approx$ 30 \%.

\subsection{Determination of $\alpha$ and $\gamma$}

\begin{figure}
%\rule{5cm}{0.2mm}\hfill\rule{5cm}{0.2mm}
%\vskip 2.5cm
%\rule{5cm}{0.2mm}\hfill\rule{5cm}{0.2mm}
\centerline{
\psfig{figure=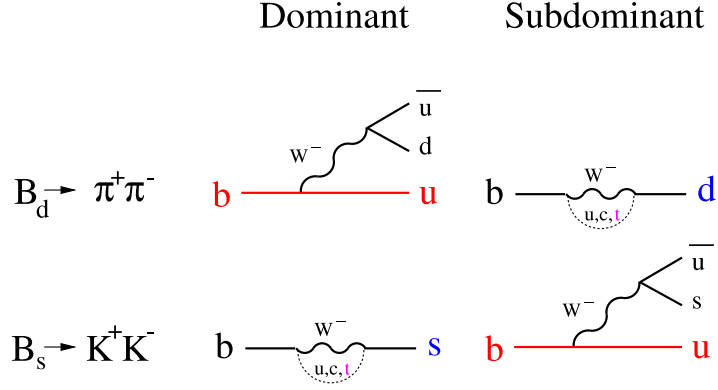,height=2.0in}}
\caption{ Left: The $x_s$ reach of BTeV
for different assumptions of the 
even yield. The curve assumes
a $S/B=1$ and $ct$ resolution of 60 fs.
Right: The $x_s$ reach of the BTeV detector. The curve 
indicates the time needed to make a 5$\sigma$ measurement.
\label{fig:peng}}
\end{figure}

The decay $B\rightarrow \pi^+\pi^-$ could have been a powerful
probe to determine $\alpha$. 
It is now well established that extracting $\alpha$ 
from $B\rightarrow \pi^+\pi^-$ is limited because
of so called " penguin pollution".
This reaction can proceed via both the tree and penguin diagrams
shown in fig. \ref {fig:peng}. The current 
experimental measurements 
$B(B^0\rightarrow K \pi) \approx 1.88 \times 10^{-5}$ 
and $B(B^0\rightarrow \pi^+ \pi^-) \approx 0.47 \times 10^{-5}$ 
indicates that the penguin amplitudes can not be neglected. 
A large number of strategies\cite{strat} have been developed to disentangle 
the penguin
and tree contributions and provide a measurement of $\alpha$.
One of the best known approaches was developed by Gronau and London\cite{gronau}.
It makes use of the isospin relation:
$$
\sqrt{2} A(B^+\rightarrow \pi^+ \pi^0)= A(B^0_d\rightarrow \pi^+ \pi^-)+
\sqrt{2} A(B^0_d\rightarrow \pi^0 \pi^0)
$$
and its CP-conjugate which form two triangles in the complex plane.
This approach relies on the determination of $B(B^0_d\rightarrow \pi^0 \pi^0)$
which is expected to be of $\mathcal{O}(10^{-6})$ and is very difficult 
to measure. Since the $\pi^0$ mesons in this decay are 
energetic, the opening angle between the two photon is small and therefore
they tend to overlap in the electromagnetic calorimeter. 

Other methods include the mode $B \rightarrow \rho \pi$ and the 
study of the CP asymmetries $B_s \rightarrow K^+K^-$ 
and $B_d \rightarrow \pi^+ \pi^-$.
The final state $\rho \pi $ offers the advantage of having a 
relatively large branching fraction. Furthermore since the $\rho$ is spin 1
while the $B$ and $\pi$ are spinless, the $\rho$ is fully polarized in this decay
which should help in maximizing the interference and therefore
achieving a better error. The BTeV collaboration following the technique proposed by 
Snyder and Quinn \cite{quinn} expects to be able to measure $\alpha$ 
with a sample of 1000-2000 background free, flavor tagged events.

The CDF collaboration plans to use the 
decay modes $B^0 \rightarrow K^+ \pi^-$,
$B_s \rightarrow K^+K^-$, $B^0 \rightarrow \pi^+ \pi^-$, 
$ B_s\rightarrow \pi^+ K^-$
and SU(3) flavor symmetry to measure $\gamma$.
Based on measured $B^0$ branching ratios and 
the production fractions $f_s$($f_d$)
for $B_s$($B_d$) we expect an event ratio of:
$$
(B^0\rightarrow K\pi): (B^0\rightarrow \pi\pi): (B_s\rightarrow KK):(B_s\rightarrow \pi K)=
4:1:2:0.5
$$
Using the latest measurement of the $B$ cross section \cite{bcross} CDF
expects between 5000 and 9000 fully reconstructed $B^0\rightarrow \pi\pi$
events in 2 fb$^{-1}$.  To study the reach in $\gamma$ CDF assumes that 
5,000 $B^0\rightarrow \pi\pi$, 20,000 $B^0\rightarrow K\pi$, 
10,000 $B_s\rightarrow KK $ and 2,500 $B_s\rightarrow \pi K $ will be reconstructed.
CDF expects to extract the physics components from the background 
by making use of the invariant mass resolution and the dE/dx information 
provided by the open cell drift chamber.
Two of the four modes of interest are self tagging  ($B^0\rightarrow 
K^{\pm}\pi^{\mp}$
and $B_s\rightarrow K^{\mp}\pi^{\pm}$) while two are CP eigenstates
($B^0\rightarrow \pi^+ \pi^-$ and $B_s \rightarrow K^+K^-$).
CDF expects to measure the time dependent CP asymmetries 
in  $B^0\rightarrow \pi^+ \pi^-$ and $B_s \rightarrow K^+K^-$ which are given by:
$$
A_{CP}= A^{dir}_{CP} cos \Delta mt + A^{mix}_{CP} sin \Delta mt 
$$
Following Fleisher \cite{fleisher} the unitarity angle $\gamma$ can been extracted
from the measured time dependent asymmetries by using the U-spin symmetry 
that relates 
$B^0\rightarrow \pi^+ \pi^-$ and $B_s \rightarrow K^+K^-$. 
In the limit of U-spin symmetry the strong phase $\theta$ and
the penguin to tree ratio $d$ in $B^0\rightarrow \pi^+ \pi^-$ and $B_s \rightarrow K^+K^-$
(denoted by $^{\prime}$) are connected by:
$$
\theta^{\prime}=\theta ~~~~~~~~~~d^{\prime}=d \frac {1-\lambda^2}{\lambda^2}
$$ 
The measured time dependent asymmetry can be used to extract 
$\beta$, $\gamma$, $d$ and $\theta$. The error in $\gamma$ varies between 6 to 15 
degrees depending to the value of $d$ which is assumed to vary between 0.1 and 0.5.

\section{$B_s$ decays}

The decay $B^0_s \rightarrow J/\psi \phi$ is the $B_s$ counterpart to the 
"golden mode" $B_d \rightarrow J/\psi K_s$. Since the 
$J/\psi$ and $\phi$ are vector mesons, the CP 
parity of the final state is a mixture of different 
CP even and odd contributions
and the CP asymmetry may be diluted by possible cancellations.
It is possible to disentangle
the CP odd and CP even contributions through an angular analysis of the 
decay products $B^0_s \rightarrow J/\psi (\ell^+ \ell^-) \phi(K^+ K^-)$.
An interesting feature is that in the standard model 
we expect CP violation in this mode 
to be small. Therefore it is a sensitive probe for 
new physics. This mode can also provide information on the mixing 
parameter $\Delta \Gamma_s$, $\Delta m_s$  and 
the weak phase $\beta_s=\arg{\left(
\frac{V_{ts}V^*_{tb}}{V_{cs}V^*_{cb}}\right) }$.  
ATLAS, CMS and LHCb have studied this decay mode extensively. 
ATLAS and CMS expect $300,000$ and $600,000$ events respectively after the 
first 3 years of operation. LHCb expects 370,000 event after 5 years
of LHC running. The result of the studies
\cite{LHC} show that $\Delta \Gamma_s $ can be measured with a precision
between 8 to 12 \% for $\Delta \Gamma_s/\Gamma_s=0.15$. The mode
appears to be sensitive to probing models of new physics such 
as the left-symmetric model with spontaneous CP violation
and the isosinglet down quark mixing model \cite{newphys}.
 
The study of the time dependent CP asymmetry in $B \rightarrow
J/\psi \eta(^{\prime})$ is another powerful test for new physics.
This decay probes the angle $\beta_s$
which is estimated to be about 0.02 in the standard model
without the cancellations that are possible 
in $B^0_s \rightarrow J/\psi \phi$. Silva 
and Wolfenstein \cite{silva} have pointed out that a measurement of
$\alpha$, $\beta$ and $\gamma$ could miss new physics.
For example if new physics arises in $B^0-\bar{B^0}$ mixing through 
a new phase $\theta^{NP}$
then a measurement of the phase in $B^0 \rightarrow J/\psi K_s$ 
will yield $2\beta^{\prime}=2\beta+\theta^{NP}$ while 
a measurement of $\alpha$ using $B^0\rightarrow \pi^+\pi^-$
(after eliminating the penguin pollution) will measure
$2\alpha^{\prime}=2\alpha-\theta^{NP}$.
In this case we would miss the new physics since 
$2\alpha^{\prime}+2\beta^{\prime}=2\alpha+2\beta$ and therefore
$\alpha^{\prime}+\beta^{\prime}+\gamma=180^\circ$.
The measurement of $\beta_s$ would instead clearly check the standard model 
since an observation of an asymmetry larger than $\mathcal{O}(\lambda^2)$ 
will be an unambiguous signal for new physics.

The decay $B \rightarrow J/\psi \eta(^{\prime})$ is very challenging
and requires excellent photon reconstruction capabilities.
The BTeV experiment will have an excellent electromagnetic
calorimeter and it expects to measure $\beta_s$ with an error $\delta\beta_s$ =0.024
after one year of running\cite{btevtdr}.

\section{Outlook and prospects}

The two asymmetric \bm-factories, PEP-II and KEKB, have achieved reliable
operation at high luminosities of a few $10^{33}{\rm cm}^{-2} {\rm s}^{-1}$ in
a remarkably short period of time after their startup. These luminosities
have enabled their experiments, \babar\ and Belle, respectively, to
observe {\it CP} violation in the decays of the $B^{o}$ meson. Operational
experience with both machines has now led to plans to 
setup super-\bm-factories operating at
$10^{35}{\rm cm}^{-2}{\rm s}^{-1}$. The hadron collider experiments 
at the Tevatron, 
CDF and D0, are beginning to produce \bm\ physics results that will 
complement the \bm-factories. Dedicated experiments at the Tevatron 
and the LHC,
BTeV, and LHCb, and the two large general
purpose experiments at the LHC, CMS and ATLAS, will begin to contribute
at very high levels of sensitivity to the study of {\it CP} violation
and rare decays in the $B$ system, starting around 2007. 
Proposal for a ``Super \bm-factory'' with a luminosity
goal of  $10^{36}{\rm cm}^{-2}{\rm s}^{-1}$ are under discussion. 

The reach of these future experiments has been studied at Snowmass \cite{shipsey}
and it summarized in table 1.
\begin{table*}
\begin{center}
\caption[]
{Comparison of {\it CP} Reach of Hadron Collider Experiments and
Super\babar. The last column is a prediction of which kind of facility
will make the dominant contribution to each physics measurement.}
\label{tab:comp_e_had_cp}
\begin{tabular}{|l|c|c|c|c|c|c|c|c|} \hline
  &CDF &D0    & BTeV       &  LHCb     &  \babar   & 10$^{35}$  & 10$^{36}$  &   \\
    & 2fb$^{-1}$ & 2fb$^{-1}$ &  10$^{7}$s & 10$^{7}$s &  Belle   & 10$^{7}$s  &  10$^{7}$s &   \\
   & &    &       &        & (2005)   &            &            &   \\ 
\hline
$\sin 2\beta$ & 0.05& 0.04&  0.011 &  0.02 &  0.037 &  0.026 &  0.008 & Equal \\
$\sin 2\alpha$& & &  0.05  &  0.05 &  0.14  &  0.1   &  0.032 & Equal \\
$\gamma\,[B_{s}(D_{s}K)]$ & $\sim$25$^{o}-45^{o}$ & & $\sim$11.5$^{o}$ &  &  &  &  & Had  \\
$\sin 2\chi$ & & &  0.024 &  0.04  &  -  &  -  &  - & Had \\
BR($B\rightarrow \pi^{o}\pi^{o}$) & & & -  &  -  & $\sim$20\% & 14 \% & 6\% &
$e^{+}e^{-}$ \\
 \hline
\end{tabular}
\end{center}
\end{table*}

It is clear that the 10$^{36}$ $e^{+}e^{-}$ machine can compete with
the hadron collider experiments on many interesting {\it CP} violating decays
and on rare decays of $B_{d}$ and $B_{u}$.
It should do better on decays involving $\tau$'s and missing
$\nu$'s since the hermeticity and energy constraints provided by running
at threshold permit one to establish the neutrino's presence in the event
by demonstrating a recoil mass consistent with zero. The experiments
at hadron collider will continue
to have better reach in the study of $B_{s}$ physics. This is a strength of
the hadron collider experiments. The  $e^{+}e^{-}$ experiments also
do not have high enough energy to study $b$-baryons or $B_{c}$ mesons.
Therefore the hadron and \bm-factories 
will be complementary and both will be needed for an exhaustive  precision
probe of the consistency of the flavor changing sector of the standard
model and in searches for New Physics.

\section{Acknowledgments}
I would like to thank the organizing committee for this conference
which brought together particle physicists and cosmologists 
in a beautiful setting.

\end{document}